# Influence of mechanical reinforcement of $MgB_2$ wires on the superconducting properties

W.Goldacker*, S.I.Schlachter

Forschungszentrum Karlsruhe, Institut für Technische Physik, P.O.Box 3640, D-76021 Karlsruhe, Germany

**Abstract**

Monofilamentary $MgB_2$-wires with a mechanical tough Nb/Cu/steel or Fe/steel sheath were prepared and characterized. The steel content was varied to investigate the reinforcement effect and the consequences for the superconducting properties of the wires, which were heat treated to achieve dense and homogeneous filaments. The use of Nb as first sheath layer, having a smaller thermal expansion coefficient than $MgB_2$, requires the application of higher amounts of steel to achieve compressive pre-stress on the filament in comparison to Fe as first wall material. With raised steel content in the sheath the critical transport currents show field dependent significant critical current and irreversibility field degradations. First $I_c$ vs. axial stress and strain experiments confirmed this observation of pre-stress induced degradations. Consequences for improved wires and for future applications will be discussed.

Keywords: $MgB_2$-wires, mechanical reinforcement, transport currents, pre-stress

*Corresponding author:
Wilfried Goldacker
Forschungszentrum Karlsruhe, Institut für Technische Physik
P.O.Box 3640, D-76021 Karlsruhe, Germany
Tel: +49-7247-824179
Fax: +49-7247-825398
e-mail: wilfried.goldacker@itp.fzk.de

## 1. Introduction

Since the discovery of superconductivity in $MgB_2$ [1], several authors reported on wires and tapes with very high transport critical current densities measured up to about $2x10^5$ $Acm^{-2}$ [2-5] and first applications of mechanically reinforced sheath composites [5]. It was found that a heat treatment of the final conductor with phase decomposition and reformation should be beneficial to the filament densification, which seems also to be supported by reinforcing the sheaths. Tapes showed significant higher $J_c$ values than wires, obviously a consequence of both a higher deformation pressure with improved phase densification during rolling and the presence of some phase texture [6]. Above a certain current level of about $2x10^5$ $Acm^{-2}$ however, transport currents could not be measured or are suppressed due to thermal instability of the conductors (large filament size, phase impurities), but extrapolations in comparison to magnetic measurements outline a potential of $J_c = 10^6$ $Acm^{-2}$ (4.2K, self field) and more. The necessary heat treatment restricts the choice of the sheath material in contact with $MgB_2$ due to chemical reasons. Ta and Nb tend out to be suitable showing minor reaction layers, while Fe does not react with decomposed $MgB_2$. For compensation of the small thermal expansion of Nb and Ta, application of a mechanical steel reinforcement is necessary. For Fe, becoming weak after annealing, it is also favourable to improve the stabilising filament pre-compression by applying steel to ensure a mechanically stable wire [6]. The goal of the presented work was a systematic investigation of the behaviour of the transport currents with varied amounts of steel reinforcement.

## 2. Experimental

Two kinds of PIT-wires were prepared with commercial $MgB_2$ powder (ALFA) and with Nb/Cu/SS and Fe/SS sheaths having three different steel contents. Most of the deformation (starting at 10 mm diameter) was done without the steel reinforcements down to approximately 1.5 mm diameter, where the outer stainless steel tubes were added, followed by further drawing steps down to 1 mm diameter. The final heat treatment was performed as given in ref. [5]. The three different applied steel tubes led to steel contents of 37, 54 and 66 % for Fe/SS sheaths (fig.1) and 33, 51, 64 % for Nb/Cu/SS sheaths (fig. 2 and tab. 1). Transport critical currents were measured with a standard 4-point method at 4.2 K and B = 0-10 T using a 1 μV/cm criterion. A first $I_c$ vs. strain experiment was performed by means of a miniature strain rig for one selected sample. The sample (40 mm) was soldered on the rig clamps. The strained sample length was 25 mm. The strain was measured via a precision strain gage clip over 15 mm sample length and the voltage taps were fixed to the sample with 10 mm spacing in between the strain gage knifes.

## 3. Results

In fig.3 the transport critical current densities for the $MgB_2$/Fe/SS wires are given. In contrast to magnetization measurements, the current density vs. field graph changes to a much smaller slope below approx. 5.5 T and consequently also to much lower critical current densities. *U(I)* curves show a sudden take off in this regime which indicates a non sufficient thermal stabilisation at such high currents. A pure Fe sheath is already expected to cause pre-compression on the filament. For increasing steel content $J_c$ decreases, the decrease becoming larger towards higher background fields. The strongest effect occurs between a pure Fe sheath and 37%, from 54 to 66% steel content the degradation effect nearly saturates. This behaviour resembles a typical pre-stress induced effect as very well known from investigations on $Nb_3Sn$ wires. As Kramer plots (Fig.4), these $J_c$ curves show clearly the degradation of $H^*$ from about 12-13 T down to about 8 T. An exact extrapolation is not possible with these data.

The critical current densities $J_c$ vs. $B$ of the $MgB_2$/Nb/Cu/SS wires show a significantly different picture (Fig.5). The $J_c$ degradation with steel amount is not as pronounced as in $MgB_2$/Fe/SS wires. The reason is that the steel has to compress the Nb layer, with the much smaller thermal contraction (see ref.[5]), too, which leads finally to a much smaller resulting amount of pre-compression in the filament. Also in these wires, non sufficient thermal stabilisation is observed below 4 T limiting the $J_c$ values, but an increased steel content obviously improves the stabilisation significantly. The highest $J_c = 65000$ $Acm^{-2}$ in self field was measured for the highest content of steel. In the moment we interpret this effect as a consequence of an improved filament densification. Regarding the corresponding Kramer plots (Fig.6) illustrates again the degradation of $H^*$ with steel content and filament pre-compression with a shift from about 11 T to 8 T.

As prove of the pre-stress induced degradation of $I_c$, a $I_c$ vs. axial strain experiment was performed on a miniature strain rig in 5 T background field for sample $MgB_2$/Fe/SS(37%) (Fig.7). The Critical current increases with applied axial strain from 14.3 A ($\varepsilon = 0\%$) to 16.2 A ($\varepsilon = 0.37\%$, $\sigma = 450$ MPa) reversible. For further strains filament cracks are formed and $I_c$ decreases irreversible, an expression of the brittleness of $MgB_2$. This result confirms a similar measurement in ref [7] on tapes in the same background field. In our case the increase of $I_c$ (13%) is about twice as large as in ref.[7] which indicates a much more pronounced pre-stress in the filament probably due to the wire geometry in our case. The strain limit of irreversible $I_c$ changes is quite similar in both experiments.

## 4. Conclusions

Stainless steel reinforcements, applied in $MgB_2$ conductors, stabilize very effectively these wires up to 450 MPa axial stress and 0.37 % axial strain. This is sufficient for all technical applications, the typical pre-stress induced $I_c$ degradations reduce upon externally applied strain. The behaviour is very similar to $Nb_3Sn$ wires and Chevrel phase wires and is consistent with the observation of a negative $dT_c/dP$ dependency in $MgB_2$ [8]. The externally applied tensile stress compensates the compressive pre-stress and leads to an increase of $J_c$. This behaviour is very favourable for application in devices like coils, where Lorentz forces and thermal stresses cause strain to the wires. The resulting increase of $J_c$ can be regarded as a mechanical reserve of the material and the device.

The observed current limitations at low or zero fields are obviously caused by thermal heating effects. This can usually be reduced by small filaments. However, the steel reinforcements led also to an improved thermal stabilisation, which is interpreted as support of the filament densification reducing porosity and the occurrence of local hot spots. Changing to multifilamentary wire concepts with small filament cross sections and improved $MgB_2$ powder quality and homogeneity is of high importance for future better stabilized wires to take advantage of the full current carrying capability in the form of real transport currents at small background fields. A systematic investigation of the pre-stress induced current degradations at different fields for all presented samples is under investigation with more precise equipment and will give much more insight in the strain dependent superconducting properties of $MgB_2$.

**Acknowledgement**

We thank H.Orschulko for preparing the wires, S.Zimmer for performing with high accuracy all measurements and H.Reiner for essential improvements of the strain rig.

**Figure captions**

Fig.1  Cross sections of $MgB_2$/Fe and three $MgB_2$/Fe/SS wires with 37, 54, 66% steel content, the wire diameter being 1 mm.

Fig.2  Cross sections of MgB2/Nb/Cu/SS wires with 33, 51, 64% steel content (1 mm diameter)

Fig.3  Transport critical current densities $J_c$ vs. $B$ for a $MgB_2$/Fe wire and three $MgB_2$/Fe/SS wires (see Fig.1).

Fig.4  Kramer plot for the $J_c(B)$ graphs of Fig.3

Fig.5  Transport critical current density $J_c$ vs. $B$ for $MgB_2$/Nb/Cu/SS wires with 33, 51, 64% steel content.

Fig.6  Kramer plot for the $J_c$ vs. $B$ graphs of Fig.5

Fig.7  $J_c$ vs. axially applied strain for a $MgB_2$/Fe/SS wire (37 % steel content)

**Table.**  Contents of the different materials in the wire composites in % of cross section

| Sample | Filament | Fe or Nb | Cu | Steel |
|---|---|---|---|---|
| $MgB_2$/Fe | 49 | 51 | - | - |
| $MgB_2$/Fe/SS25 | 30 | 33 | - | 37 |
| $MgB_2$/Fe/SS35 | 22 | 24 | - | 54 |
| $MgB_2$/Fe/SS45 | 16 | 18 | - | 66 |
| $MgB_2$/Nb/Cu/SS25 | 28 | 18 | 21 | 33 |
| $MgB_2$/Nb/Cu/SS35 | 20 | 13.5 | 15.5 | 51 |
| $MgB_2$/Nb/Cu/SS45 | 14 | 10 | 12 | 64 |

Table

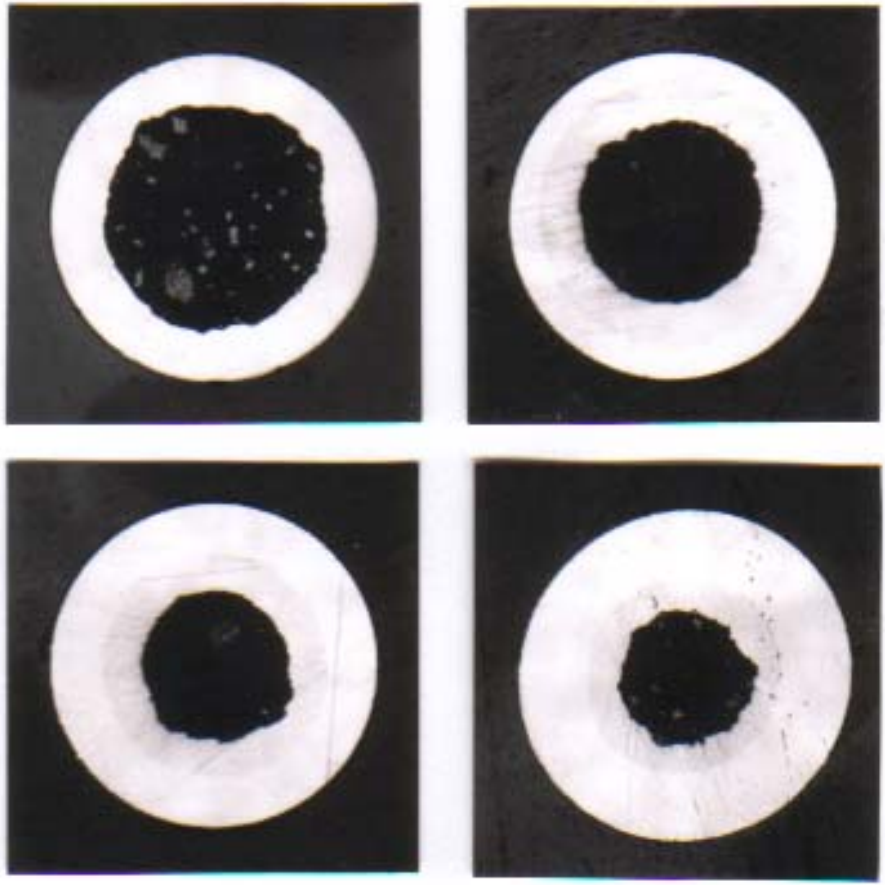

Fig.1.

Goldacker et al.

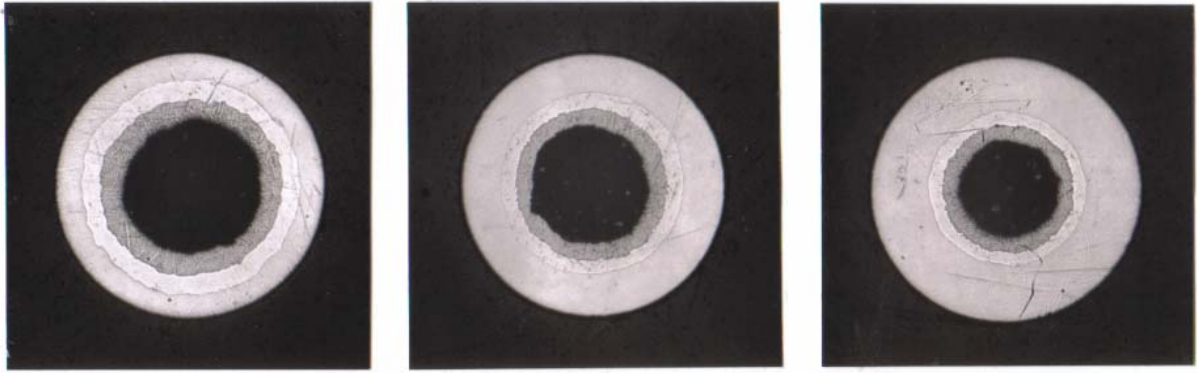

Fig. 2

Goldacker et al.

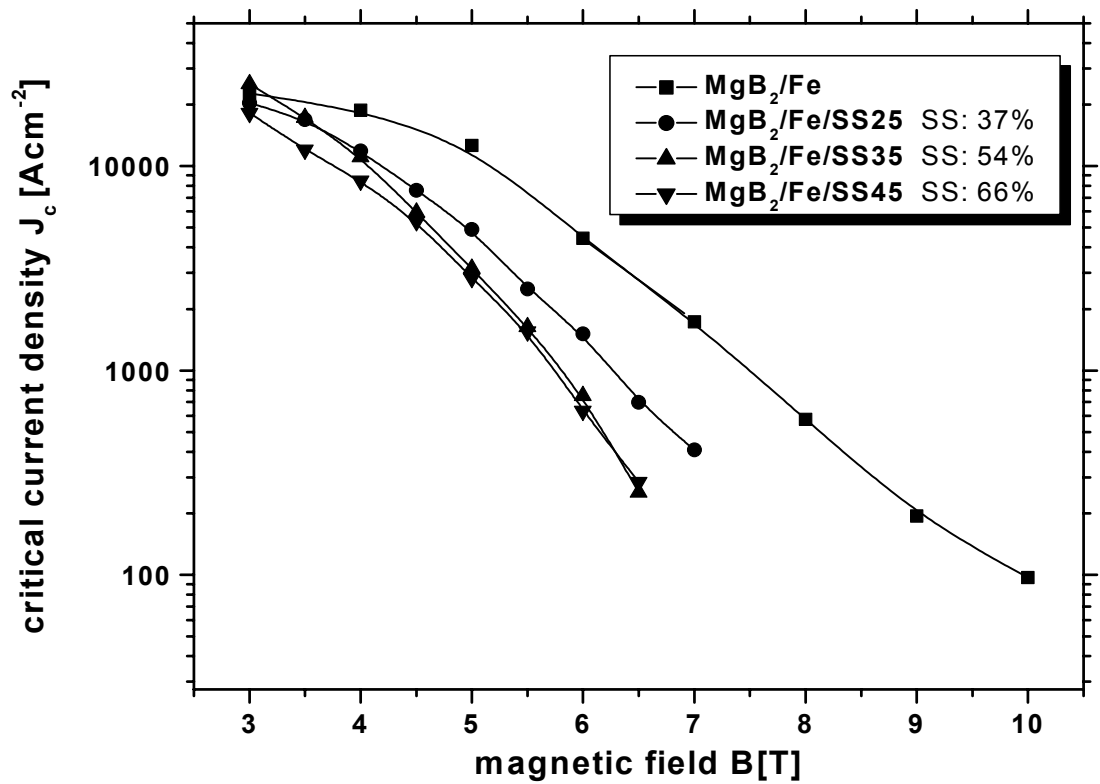

Fig. 3

Goldacker et al.

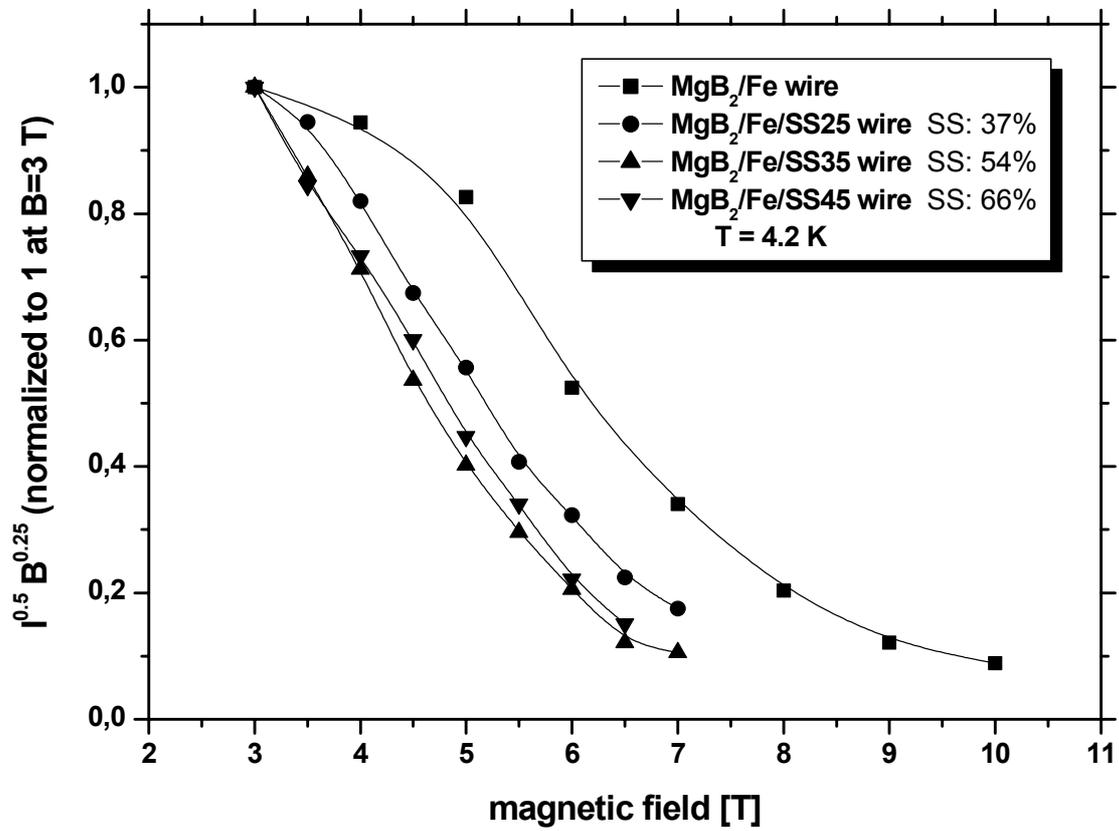

Fig. 4

Goldacker et al.

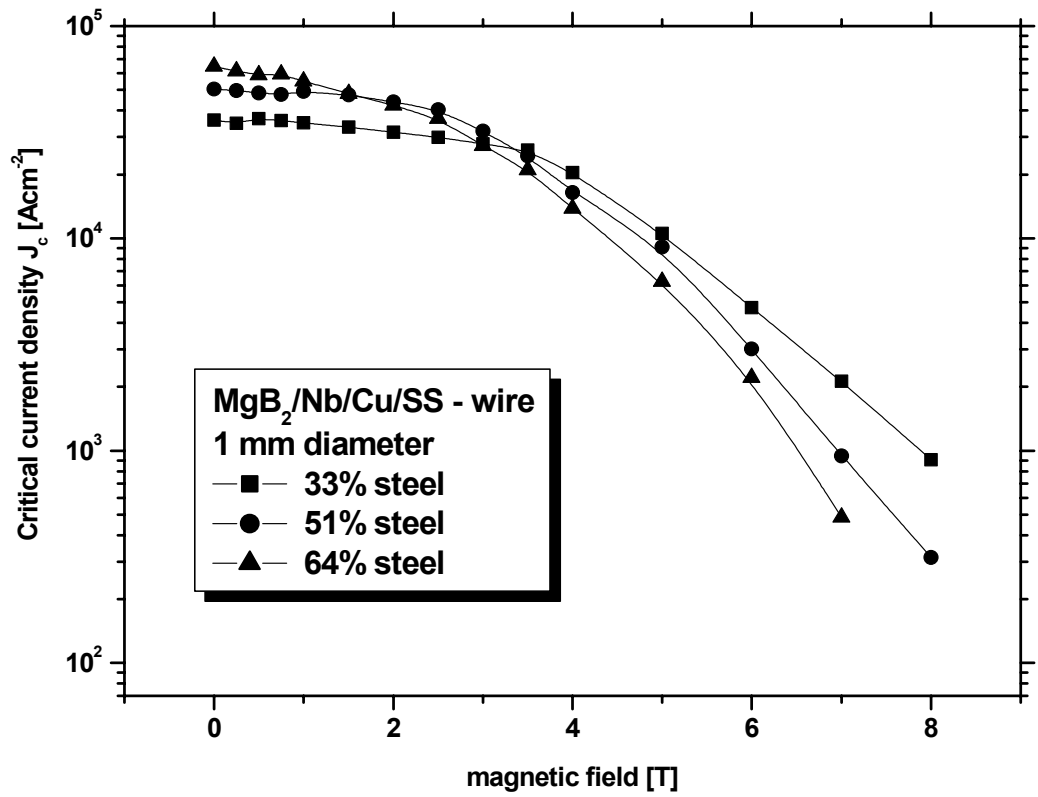

Fig. 5

Goldacker et al.

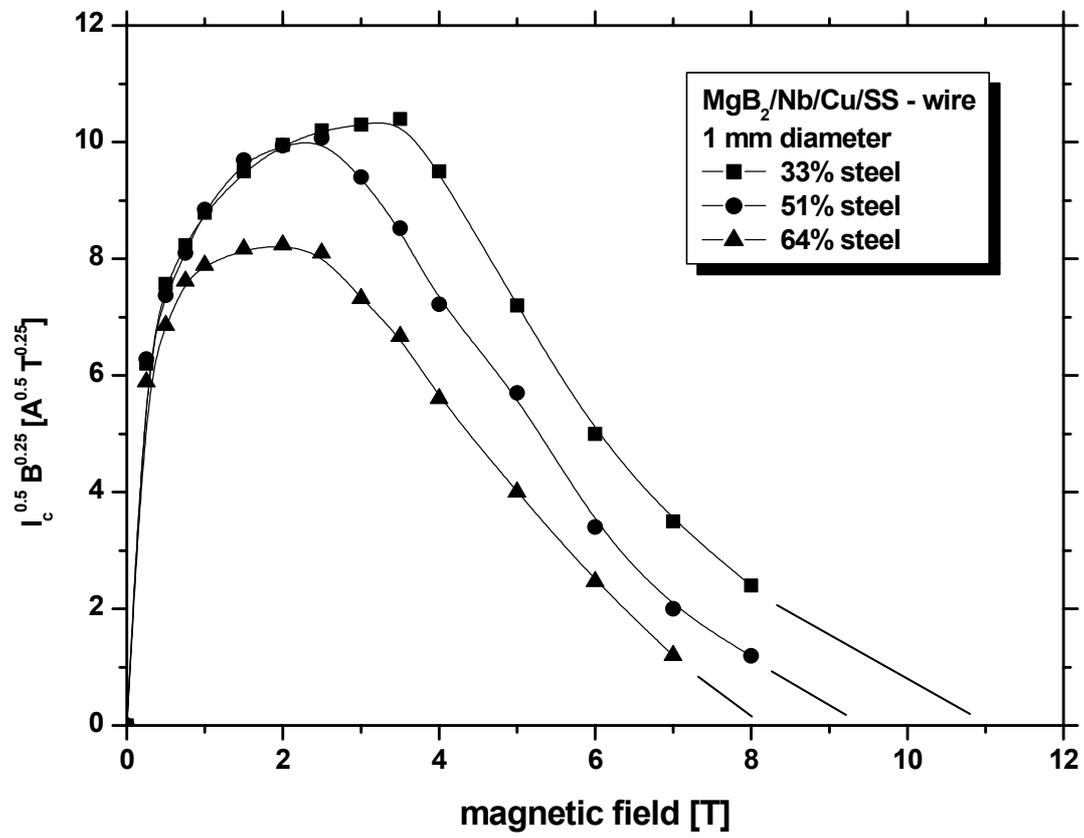

Fig. 6

Goldacker et al.

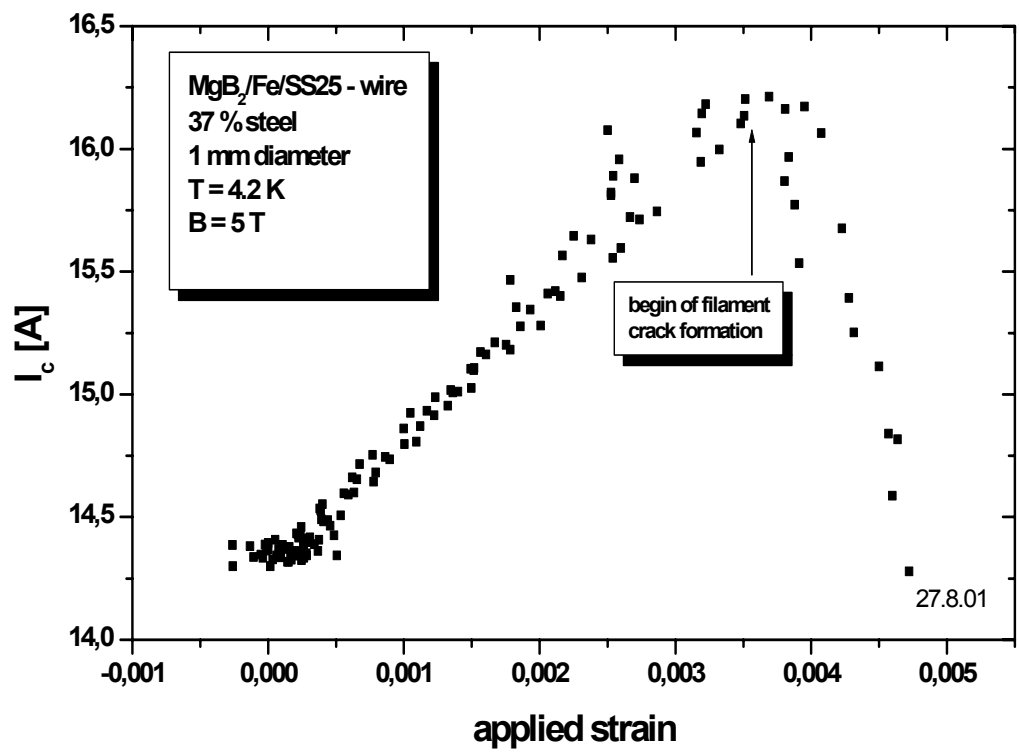

Fig. 7

Goldacker et al.